# Drop-on-demand printed negative dielectric anisotropy liquid crystal droplets for adaptive complex beam manipulation and assessment

Jinge Guo[1,†], Xuke Qiu[1,†], Runchen Zhang[1], Qihao Han[1], Liangyu Deng[1], Yishun Lu[2], Zimo Zhao[1], Mengmeng Li[1], Waqas Kamal[1], Junseok Ma[3], Yongge Ma[4], Steve J Elston[1], Alfonso A. Castrejón-Pita[1], Stephen M Morris[1,*], and Chao He[1,*]

*[1]Department of Engineering Science, University of Oxford, Parks Road, Oxford, OX1 3PJ, UK*

*[2]Oxford e-Research Centre, University of Oxford, 7 Keble Rd, Oxford OX1 3QG*

*[3]Nature Sciences Research Institute, Korea Advanced Institute of Science and Technology (KAIST), 291 Daehak-ro, Yuseong-gu, Daejeon 34141, Republic of Korea*

*[4]School of Computer Science, Peking University, No. 5 Yiheyuan Road, Haidian District, Beijing, P.R.China*

*†These authors contributed equally to this work*

**Corresponding authors: stephen.morris@eng.ox.ac.uk; chao.he@eng.ox.ac.uk*

## Abstract

**Adaptive manipulation of vectorial optical fields are important for optical metrology, imaging, and structured light related applications, yet existing approaches often rely on bulky or sequentially operated systems. Here we demonstrate an inkjet-printed negative dielectric anisotropy nematic liquid crystal droplet platform that unifies adaptive complex beam generation and full-vectorial optical field sensing within a single printed architecture. For complex beam generation, voltage-driven director reconfiguration in the droplets produces tunable birefringence and wavelength-dependent polarization textures, including skyrmionic-like optical fields. For adaptive full-vectorial optical field sensing, the same droplet array enables spectral and polarization retrieval through wavelength-dependent intensity patterns and division-of-wavefront polarimetry, while also functioning as a microlens array for Shack–Hartmann wavefront sensing to reconstruct phase. These results establish negative dielectric anisotropy liquid crystal droplets as a scalable soft-matter photonic system for adaptive beam manipulation and multidimensional optical field characterization.**

## Introduction

Liquid crystals (LCs) are widely recognized for their anisotropic optical properties and electric field-responsive molecular alignment, making them attractive for a broad range of photonic and optoelectronic applications. Conventional LC devices typically rely on sandwich-glass cell architectures that require precise cell-gap control (the separation between the glass substrates) and carefully prepared alignment layers. Recent advances in inkjet printing have introduced a digital, maskless, and additive route for fabricating LC-based optical elements[1-5]. This approach enables precise control over droplet volume and placement on the substrate, compatibility with a diverse range of substrates, including those that are conformable. Moreover, inkjet-printed LC architectures offer significant advantages in scalability, material efficiency, and fabrication simplicity, opening new opportunities for printed micro-optics[6], flexible photonic devices[1], and integrated optical systems[7, 8].

A particularly attractive technological application of printed LC droplets is in the characterization of complex of optical fields. The accurate characterization of vectorial optical fields,  is essential for a diverse range of applications such as adaptive optics[9], structured light manipulation[10], biomedical imaging[11], and optical communications[12]. However, the majority of existing vectorial sensing techniques rely on bulky optical assemblies or sequential measurement schemes that require multiple optical components or repeated exposures. These constraints limit their compactness, scalability, and real-time applicability. Recently, inkjet-printed LC droplet arrays have emerged as a promising soft-matter platform for compact optical manipulation and characterization [13, 14]. By exploiting the spatially varying birefringence within each LC droplet, such systems can encode both the polarization and phase information across the transmitted optical field, enabling single-shot retrieval of multiple optical parameters within a compact device architecture.

Despite these advances, several fundamental limitations remain with regards to the recent demonstration of printed droplet-based sensing platforms which have employed the use of positive dielectric anisotropy nematic LC droplets, in both optical field sensing and generation. First, as the director aligns homeotropically in these droplets, the optical response is predominantly passive[13], meaning that the intrinsic electro-optic tunability of LC is not fully exploited for either sensing or active beam generation, unless in-plane electrode geometries are used[15], which in turn can result in an undesirably non-uniform electric field and distorted director profile[16]. Second, the use of positive dielectric anisotropy LC droplets hinders the realization of wavelength-dependent complex beam generation and integrated spectral sensing[13], as their birefringent response is largely static and exhibits limited wavelength-dependent tunability. Third, polarization retrieval based on positive dielectric anisotropy nematic LC droplets is highly sensitive to the droplet boundary profile[13], imposing stringent requirements on printing uniformity and fabrication conditions.

Nematic LCs with negative dielectric anisotropy, on the other hand, offer a promising route to overcome the limitations mentioned above. In such nematic LCs, the director reorients perpendicular to the direction of the applied electric field, enabling electrically tunable birefringence and more flexible control over the internal optical anisotropy. Here, we demonstrate the use of inkjet-printed droplets of a negative dielectric anisotropy nematic LC (LC-VATS14, $\Delta\varepsilon = -2.72$) as a multifunctional soft-matter photonic platform. The printed droplet array functions as a compact adaptive optical sensor capable of retrieving multiple parameters of an optical field within a single architecture. By exploiting the spatially varying birefringence and microlensing properties of the droplets, the system enables measurement of intensity, polarization, phase, and wavelength. Interestingly, we further find that the same droplets can also act as electrically tunable generators of complex vectorial optical fields. In particular, voltage-driven reconfiguration of the internal director structure produces a rich tapestry of polarization textures, including voltage- and wavelength-dependent skyrmionic polarization fields. Collectively, these results establish printed negative dielectric anisotropy nematic LC droplets as an adaptive photonic platform that integrates compact vectorial field sensing with adaptive complex beam generation within a single scalable architecture.

## Results

### 1. LC droplet characteristics

Unlike conventional positive dielectric anisotropy nematic LCs, negative dielectric anisotropy nematic LCs reorient perpendicularly to the orientation of the applied electric field[17, 18]. By leveraging this property, we have fabricated arrays of inkjet-printed negative dielectric anisotropy LC (VATS14, Instec Inc.) droplets that undergo voltage-tunable director reconfiguration. As shown in Fig. 1a,b, arrays of negative dielectric anisotropy nematic LC droplets are fabricated via inkjet printing onto ITO-coated substrates, forming well-defined hemispherical microstructures with controlled size and spacing, and sandwiched between two electrodes to enable the application of an out-of-plane AC electric field. Under polarized optical microscopy (POM, where two orthogonally oriented polarizers are placed in the illumination and detection paths), the droplets exhibit a transition from a bend-dominated to twist-favored director profiles, producing spiral-like director fields and distinctive interference fringes (**Fig. 1**c–f). These twist-driven structures arise from the relative magnitudes of the elastic constants of negative dielectric anisotropy LCs ($K_2 < K_1, K_3$) [19]and represent a symmetry-broken configuration that spontaneously forms under an external bias (see **Fig. S1** for details).

The voltage-controlled director response has three key implications. **1)** The application of an out-of-plane electric field in these printed negative dielectric anisotropy nematic LCs induces strong reorientation of the director toward the plane of the substrates, significantly enhancing the birefringence and enabling full $0 - \pi$ retardance even in sub-100 μm diameter droplets. This voltage-driven retardance modulation allows the droplets to achieve high phase contrast without requiring large, printed LC volumes, ensuring effective optical functionality within a microscale footprint (See Mueller matrix details in **Fig. S2**). **2)** The spiral morphologies generate wavelength-dependent birefringence, as evidenced by the distinct contrast between broadband and red-filtered POM images (Fig. 1d–e). These reproducible spectral variations can be quantitatively correlated with director reconfiguration, thereby providing a natural transduction mechanism for multispectral or wavelength-selective sensing, which is discussed further later in this article. **3)** The twist-dominated fields naturally translate into topologically nontrivial polarization textures (schematically illustrated in Fig. 1c), allowing negative dielectric anisotropy LC droplets to serve as compact and reconfigurable elements for structured-light generation.

Together, these features indicate that negative dielectric anisotropy nematic LC droplets not only offer a robust micro-optical platform but can also act as an effective bridge connecting director profile configurations with macroscopic photonic functionality. Specifically, the field-driven transition from axisymmetric to twisted director configurations establishes an optical environment that facilitates the emergence of Skyrmionic-like polarization textures, as explored in the next section.

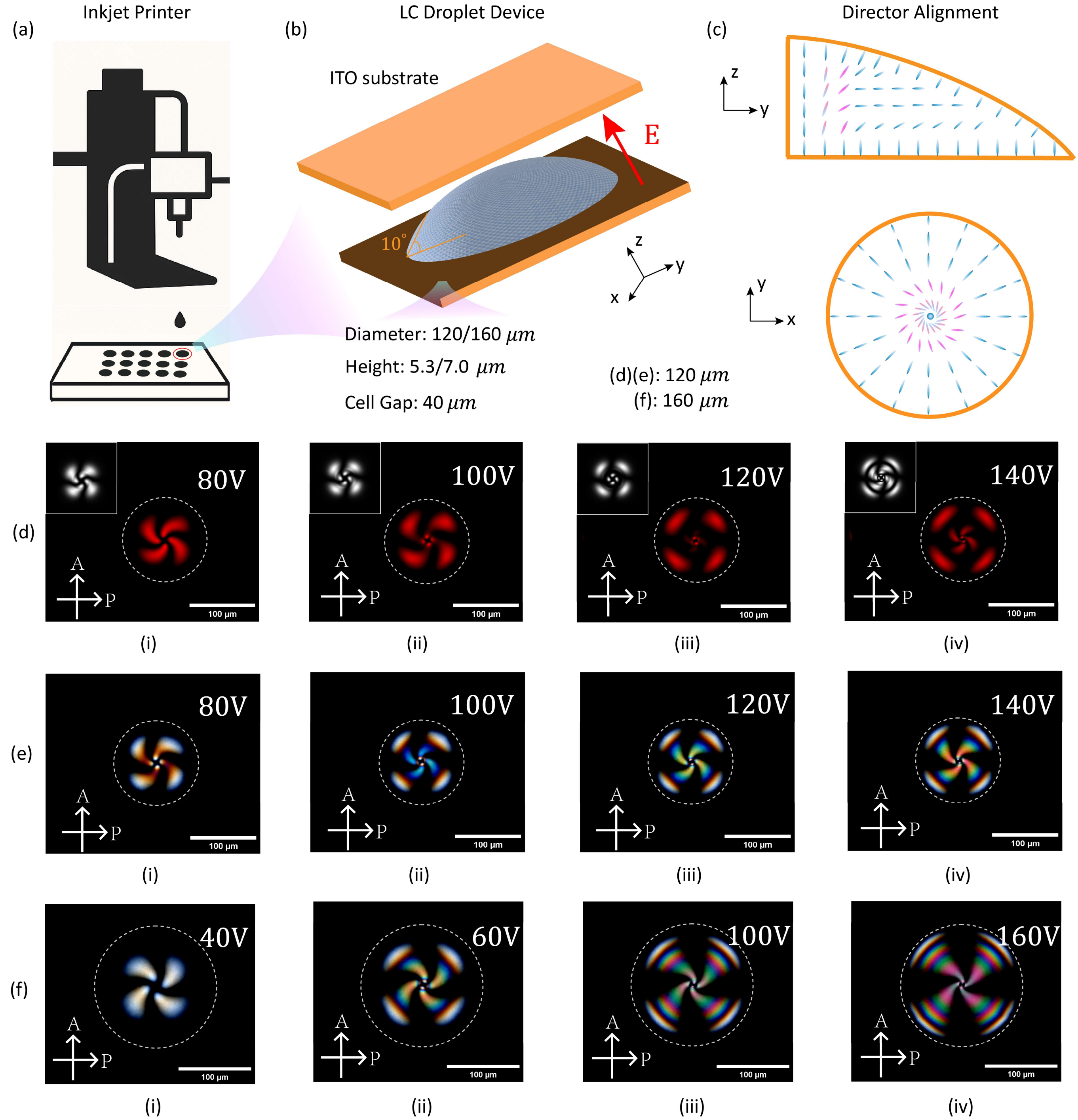

**Fig. 1 Schematic overview of the Inkjet-printed negative dielectric anisotropy nematic LC (VATS14) droplets.** (a): Illustrations of the Inkjet printer and printed LC droplets on an ITO-coated homeotropic aligned glass substrate. (b): A schematic illustration of printed negative dielectric anisotropy LC droplets with diameter around 120 μm, contact angle of ≈10° (exaggerated in Figure 1(b)), printed on an ITO-coated glass substrate with a homeotropic alignment layer and then a second ITO-coated glass substrate attached on top with 40 μm spacers. A 1kHz AC voltage is applied across the two ITO glass substrates. (c): Illustrations of Y-Z and X-Y projections of the director alignment of printed negative dielectric anisotropy nematic LC droplets when subjected to an out-of-plane electric field. POM images of printed negative dielectric anisotropy nematic LC droplets (diameter 120 μm) subjected to different applied voltages at a frequency of 1 kHz, viewed with (d) and without (e) a red filter on the microscope. (d) inserts: Simulated POM images of the negative dielectric anisotropy nematic LC droplets for different applied voltages. (f) POM images of the LC droplets with larger diameters (160 μm) for different applied voltages. The single headed black arrows denote the orientations of the polarizer and analyzer. These images were obtained at a temperature of 25°C.

## 2. Tunable Skyrmionic field generation

As evidenced in the previous section, the application of an external electric field induces spiral director fields in the printed negative dielectric anisotropy nematic LC droplets, transforming the initially axisymmetric configuration into twist profiles. This voltage-driven reorientation of the optic axis gives rise to spatially rotating polarization patterns that act as precursors to skyrmionic Stokes-field configurations. Analogous to magnetic skyrmions, stable nanoscale quasiparticles with topological protection and low energy requirements for manipulation[20, 21], their optical counterparts manifest as polarization textures with nontrivial topology[22-28], characterized by an integer skyrmion number $N_{sk}$. In an order-2 skyrmion (**Fig. 2a**), for instance, the polarization vector completes two full rotations across the beam cross-section.

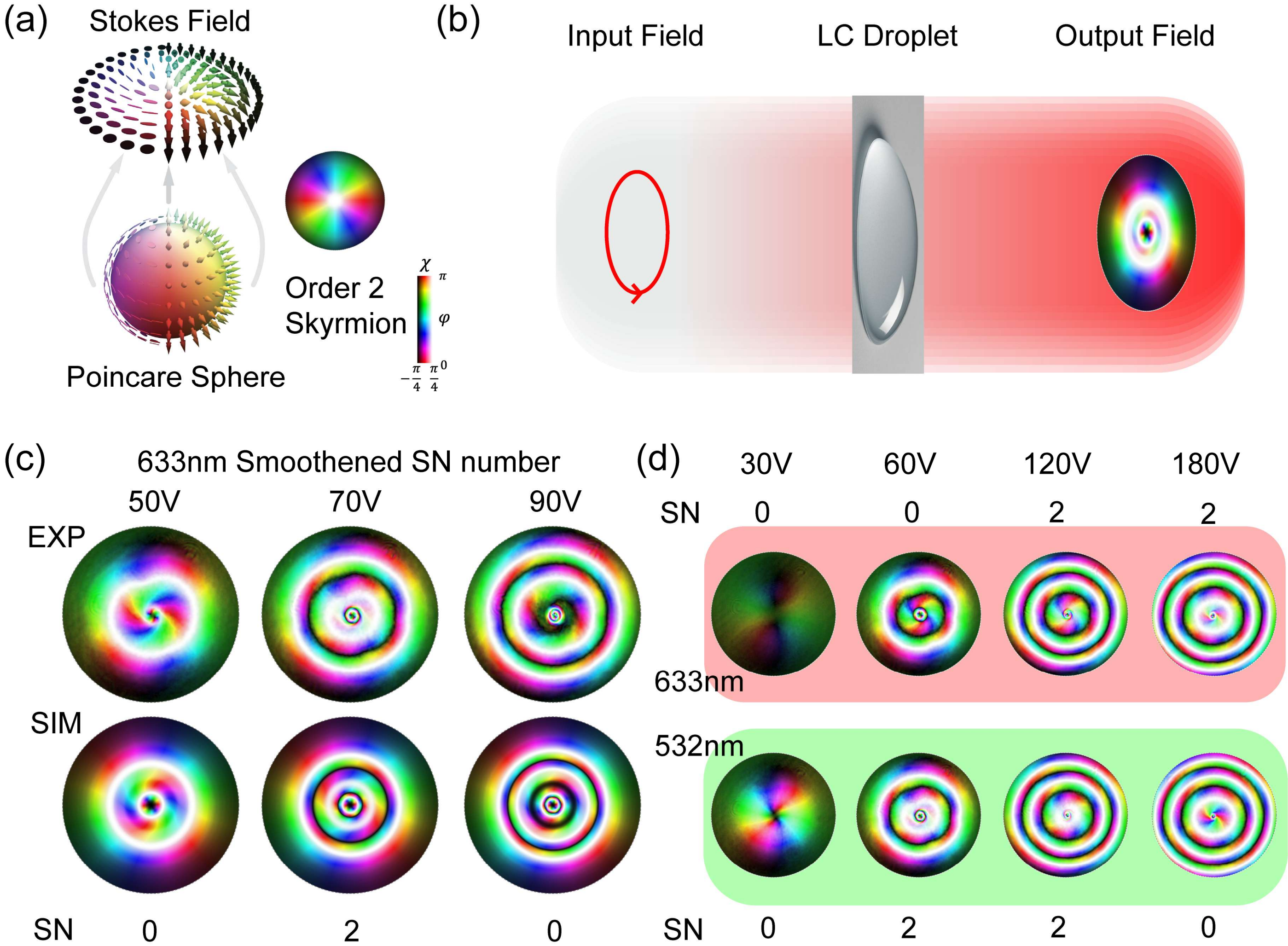

**Fig. 2 Adaptive optical skyrmion generation using printed negative dielectric anisotropy nematic LC droplets with a diameter of 160 µm on the glass substrate.** (a) Mapping between experimentally reconstructed Stokes field distributions and their representation on the Poincaré sphere, highlighting an order-2 skyrmion structure. (b) Schematic illustrating complex polarization field generation under left-circularly polarized illumination through a negative dielectric anisotropy nematic LC droplet. (c) Experimental and simulated polarization fields at different voltages (50, 70, 90 V), showing electric field-driven evolution of topological textures and corresponding changes in skyrmion number $N_{sk}$. (d) Wavelength-dependent and adaptive topological switching of $N_{sk}$ under dual-wavelength excitation (633 nm and 532 nm), demonstrating voltage-controllable and wavelength-enabled skyrmion encoding.

To explore the emergence of such structures in our system, we illuminated the printed negative dielectric anisotropy nematic LC droplets with left-circularly polarized light and reconstructed the transmitted Stokes fields under varying voltage bias conditions (Fig. 2b-d). The observed voltage-dependent transition from axisymmetric to twisted director configurations demonstrates how electric-field-driven director reorientation naturally gives rise to skyrmionic-like textures.

To quantify the emergence of these topological configurations, we computed the skyrmion number $N_{sk}$ from the experimentally reconstructed Stokes-vector fields using the standard skyrmion integral (see **Supplementary Note 2** for details). In doing so, we use engineering solutions to characterize the topological information of the generated field. Although increasing the voltage may not break down the intrinsic topology of the beam, our focus here is on the generation of distinct topological information in the resulting optical field, as characterized using the engineering solutions described in **Supplementary Note 2**. **1)** Under increasing bias, the Stokes-field configuration undergoes discrete transitions between $N_{sk} = 0$ to $N_{sk} = 2$, confirming that voltage-driven twist in negative dielectric nematic LC droplets enables electric-field-tunable skyrmion generation. **2)** Under dual-wavelength illumination at 633 nm and 532 nm, the two spectral channels display distinct evolution pathways of $N_{sk}$ under identical voltage bias conditions, revealing wavelength-enabled control arising from dispersion in the retardance. **3)** When both parameters vary simultaneously, the red and green channels exhibit cross-over transitions of $N_{sk}$: at 30 V both (0, 0); at 60 V (0, 2); at 120 V (2, 0); and at 180 V (0, 0), demonstrating adaptive switching of topological states mediated by coupled electro-optic and spectral effects.

Collectively, these observations establish the printed negative dielectric anisotropy nematic LC droplet system as a compact, electrically and spectrally tunable optical skyrmionic field generator, in which dual-parameter control enables programmable transitions between distinct integer-valued topological states. This adaptive tunability provides the physical basis for spectral–topological multiplexing, potentially motivating further applications in optical skyrmion communication. A potential framework is outlined in Supplementary Note 2.

### 3. Adaptive full-vectorial field optical sensing

The capability of negative dielectric anisotropy nematic LC droplets to generate complex vectorial light fields naturally suggests their complementary role in optical field characterization. Accurate and real-time characterization of optical fields, particularly the retrieval of polarization and phase, is essential for applications such as biomedical diagnostics[29-38], adaptive optics[39, 40], and vectorial beam characterization[41-45]. However, most existing sensing systems rely on bulky optical assemblies, complex interferometric configurations, or sequential measurement schemes that limit compactness and real-time applicability[46-54]. Printed LC droplet arrays have recently emerged as a promising compact alternative. In particular, droplets based on positive dielectric anisotropy nematic mixtures (e.g., E7) have been demonstrated for reconstruction of local Stokes parameters and wavefront phase[13], where each droplet functions as both a polarization-sensitive element and a microlens.

In the following, we demonstrate that printed negative dielectric anisotropy nematic LC droplets can offer enhanced functionality as an adaptive optical sensing platform that can retrieve multiple parameters of the optical field. The system operates under two electrical states (field ON and OFF), enabling different sensing modalities within a unified architecture. Under the electric field ON state, the intrinsic spectral response of the droplets enables wavelength sensing through spectral encoding, while spatially varying birefringence supports polarization sensing via division-of-wavefront polarimetry. When the electric field is switched OFF, the droplets behave as microlenses, enabling wavefront reconstruction through a Shack–Hartmann analysis. Compared with previously reported positive dielectric anisotropy nematic LC droplet platforms, the negative dielectric anisotropy LC droplets exhibit enhanced spectral sensitivity, while also providing improved robustness and sensing accuracy. Together, these results establish printed negative dielectric anisotropy LC droplets as an adaptive sensing architecture that integrates spectral, polarization, and phase characterization within a single compact and reconfigurable platform.

### 3.1 A full-vectorial field optical sensing framework

To enable structured reconstruction of the optical field, we implement a reconfigurable sensing framework based on printed negative dielectric anisotropy LC droplet arrays, as illustrated in **Fig. 3a**. The different sensing functionalities arise from the distinct optical responses of the LC droplets under different electric field conditions, together with a fixed polarization analysis configuration consisting of external polarizers.

For the electric field ON state, the LC director reorients along the applied field, increasing the effective birefringence and forming spatially varying retardance distributions within each droplet. When combined with the external polarizer–analyzer configuration, this spatially varying birefringence is converted into measurable intensity patterns. These intensity distributions exhibit wavelength-dependent polarization textures, enabling spectral encoding (Mode 1), while simultaneously providing the spatial modulation required for polarization analysis via division-of-wavefront polarimetry (Mode 2).

When the electric field is switched OFF, the local LC director relaxes to the equilibrium configuration, and the droplet curvature dominates the optical response. In this state, each droplet acts as a microlens, converting local wavefront gradients into focal spot displacements, which are then used for phase reconstruction via a Shack–Hartmann sensing (Mode 3). In this mode, the measurement relies on geometric focusing rather than polarization analysis. Combined, these distinct operating regimes allow the same droplet array to support a range of multiple sensing modalities within a single device, as summarized in **Fig. 3a**.

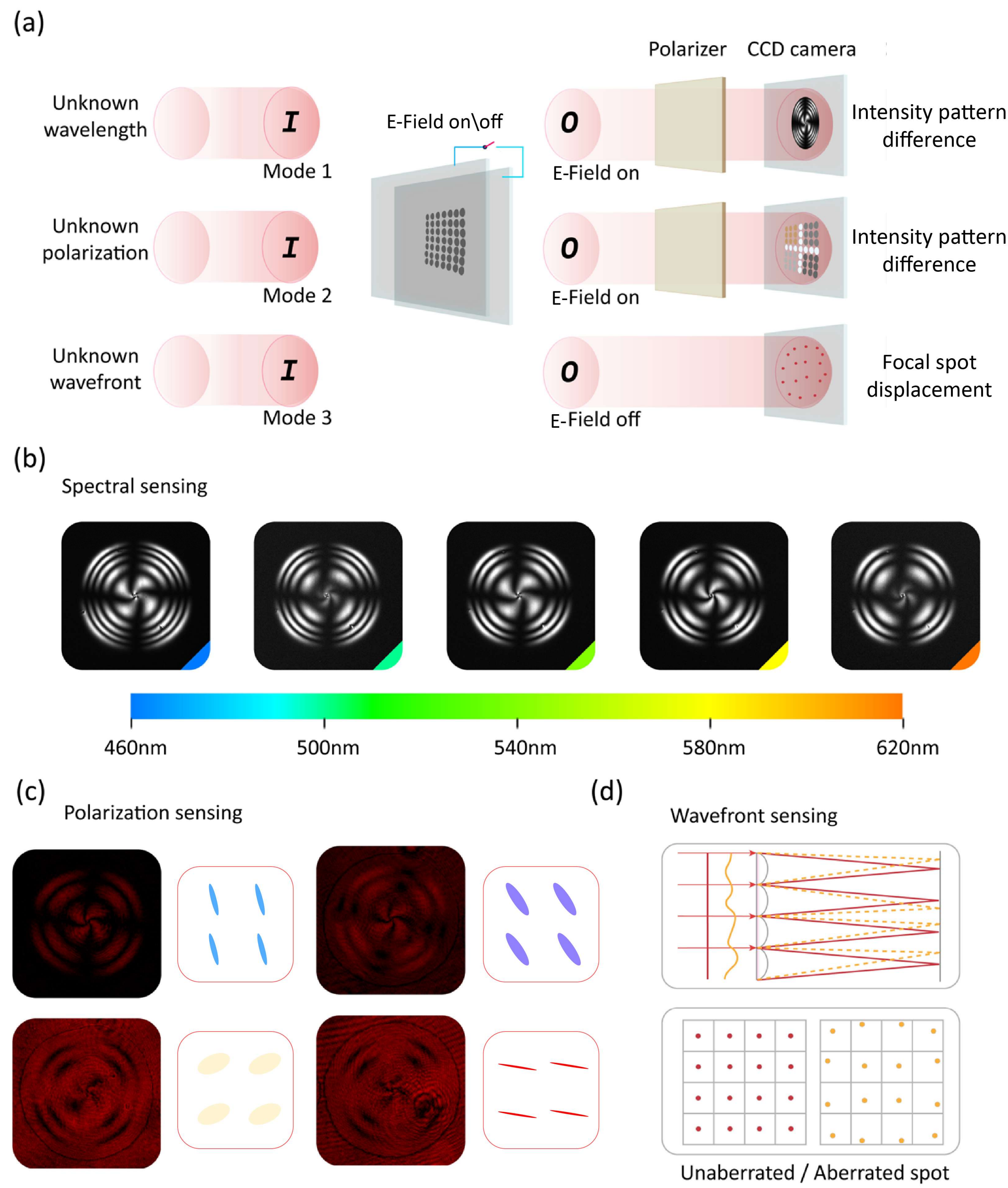

**Fig. 3 Structured full-vectorial optical sensing framework using printed negative dielectric anisotropy nematic LC droplet arrays.** (a) Conceptual illustration of the different forms of sensing. The printed LC droplet array operates under two different electric field conditions (field ON and OFF), enabling three sensing modalities for reconstructing different parameters of the incident optical field. For the field ON state, the droplets act as spatially varying birefringent elements, allowing (Mode 1) spectral sensing via wavelength-dependent intensity patterns using crossed polarizers, and (Mode 2) polarization sensing via division-of-wavefront polarimetry. For the field OFF state, the droplets behave as

microlenses, enabling (Mode 3) phase sensing through Shack–Hartmann wavefront reconstruction. This unified yet reconfigurable platform allows decomposition of the optical field into wavelength, polarization, and phase components. (b) Spectral sensing demonstration. Representative POM images of a single negative dielectric anisotropy nematic LC droplet with an electric field applied, recorded under monochromatic illumination at different wavelengths across the visible range. The images are captured using a monochrome camera under a fixed polarizer–analyzer configuration. The wavelength-dependent birefringent retardance produces distinct intensity patterns, enabling spectral encoding. (c) Representative polarization sensing. Left: measured intensity patterns within each droplet under different incident states of polarization. Right: reconstructed polarization states represented as polarization ellipses. The full Stokes parameters were retrieved from the spatially resolved intensity values inside each droplet through calibrated inversion; the ellipse color and ellipticity encode the polarization azimuth and handedness, respectively. (d) Phase sensing principle based on Shack–Hartmann wavefront sensing. Aberrated wavefronts produce displaced focal spots compared with the reference (unaberrated) case, allowing phase gradients to be extracted from spot displacements.

All sensing modalities presented in this work are implemented using nematic LC droplets with a uniform diameter at the glass substrate of 160 μm. This size is selected as a compromise between sufficient birefringent modulation for polarization and spectral encoding, and adequate microlens performance for wavefront sensing. A detailed discussion of the size optimization is provided in **Supplementary Note 1**.

### 3.2 Spectral sensing (Mode 1)

We first demonstrate the spectral sensing functionality (Mode 1) under the electric field ON state, as introduced in Fig. 3b. In this configuration, the spatially varying birefringence of each nematic LC droplet, combined with the fixed polarizer–analyzer arrangement, converts wavelength-dependent retardance into characteristic intensity patterns. These patterns serve as a unique optical signature of the incident wavelength, enabling spectral encoding within a single droplet.

To quantify this capability, monochromatic illumination spanning the visible range from 460 nm to 619 nm was applied with a step size of 1 nm, resulting in 160 distinct wavelength conditions. For each wavelength, the transmitted POM image of a single nematic LC droplet was recorded using a monochrome camera. The resulting dataset captures the wavelength-dependent evolution of the droplet intensity patterns.

A regression model was then employed to map the measured intensity patterns to the corresponding wavelengths. A subset of 20 images was used for training, while the remaining 140 images were used for evaluation. Details of the model architecture and training procedure are provided in **Supplementary Information Note 3**. The reconstructed versus reference wavelengths are shown in **Fig. 4a**. Specifically, the left panel plots the predicted wavelength as a function of the ground-truth value, demonstrating a near-linear relationship across the entire spectral range (460–619 nm), with data points closely distributed along the diagonal. The right panel presents the distribution of reconstruction residuals, which is narrowly centered around zero with a standard deviation of ~1.5 nm, indicating both high accuracy and stability of the spectral reconstruction. The model achieves a mean absolute error (MAE) of 1.22 nm and a root mean square error (RMSE) of 1.48 nm, with a near-zero mean residual, indicating high spectral resolution and stability.

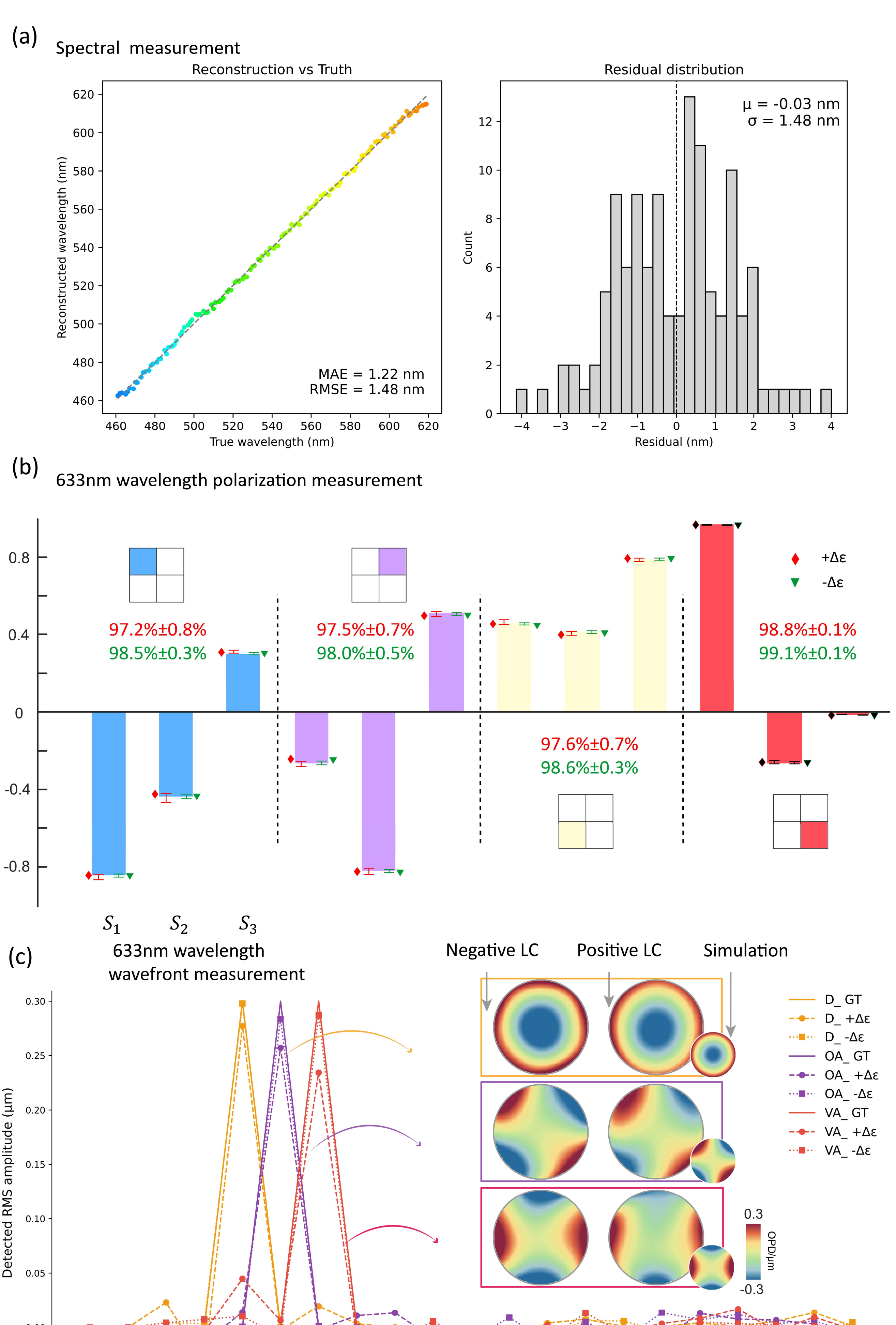

(a)
Spectral measurement
Reconstruction vs Truth
Reconstructed wavelength (nm)
True wavelength (nm)
MAE = 1.22 nm
RMSE = 1.48 nm
Residual distribution
Count
Residual (nm)
μ = -0.03 nm
σ = 1.48 nm
(b)
633nm wavelength polarization measurement
97.2%±0.8%
98.5%±0.3%
97.5%±0.7%
98.0%±0.5%
97.6%±0.7%
98.6%±0.3%
98.8%±0.1%
99.1%±0.1%
+Δε
-Δε
$S_1$
$S_2$
$S_3$
(c)
633nm wavelength
wavefront measurement
Negative LC
Positive LC
Simulation
Detected RMS amplitude (μm)
Zernike Mode Index
D_ GT
D_ +Δε
D_ -Δε
OA_ GT
OA_ +Δε
OA_ -Δε
VA_ GT
VA_ +Δε
VA_ -Δε
OPD/μm
0.3
-0.3

**Fig. 4 Experimental demonstration of multimodal optical field sensing.** (a) Spectral sensing performance (Mode 1). Left: reconstructed wavelength versus reference wavelength, showing high linearity across the visible range. Right: distribution of reconstruction residuals, with a mean error close to zero and a standard deviation of ~1.5 nm, indicating high spectral resolution. (b) Polarization sensing performance (Mode 2) at 633 nm. Retrieved normalized Stokes parameters ($S_1$, $S_2$, $S_3$) for multiple input polarization states, compared with reference values. Results from negative dielectric anisotropy LC droplets (VATS14, Instec Inc.) and positive dielectric anisotropy LC droplets (E7, Synthon Chemicals Ltd.) are shown, demonstrating improved accuracy and reduced deviation for the negative dielectric anisotropy nematic LC platform. (c) Phase sensing performance (Mode 3) at 633 nm. Left: reconstructed Zernike coefficients for different aberration modes. Right: comparison of reconstructed wavefronts for negative dielectric anisotropy LC, positive dielectric anisotropy LC, and simulation. The negative dielectric anisotropy LC droplets show improved reconstruction fidelity and reduced error.

### 3.3 Polarization sensing (Mode 2)

We next demonstrate polarization sensing (Mode 2) under the electric field ON state. When an AC electric field is applied across the droplet (between the glass substrates), the local LC director reorients perpendicular to the field direction, producing spatially varying birefringence within each droplet. As for the spectral sensing, the fixed polarizer–analyzer configuration ensures that the spatially varying retardance converts the incident state of polarization into characteristic intensity distributions across the droplet. These intensity patterns are then used to retrieve the full Stokes vector ($S_0$–$S_3$) via a pre-calibrated instrument matrix, enabling single-shot division-of-wavefront polarimetry. The corresponding polarization sensing performance is shown in Fig. 4b. In this panel, the retrieved normalized Stokes parameters (S1, S2, S3) for multiple input polarization states are plotted alongside their reference values. The close overlap between measured and reference data confirms accurate reconstruction of the polarization state. The reconstructed normalized Stokes parameters ($S_1$, $S_2$, $S_3$) for multiple input polarization states show strong agreement with the reference values, with deviations within a few percent.

A comparison of printed negative dielectric anisotropy LC droplets (VATS14) and positive dielectric anisotropy LC droplets (E7) shows that the negative dielectric anisotropy LC platform consistently provides improved reconstruction accuracy and reduced variation. These results confirm that the spatially resolved intensity patterns within each droplet provide sufficient information for reliable polarization reconstruction.

### 3.4 Wavefront sensing (Mode 3)

Finally, we demonstrate wavefront sensing (Mode 3) under the electric field OFF state. When the electric field is removed, the LC director relaxes to their equilibrium configuration, and the droplet curvature dominates the optical response. In this state, each droplet effectively functions as a microlens, converting local wavefront gradients into focal spot displacements in a Shack–Hartmann configuration. The local wavefront slopes were extracted from the displacement of each focal spot relative to a reference, and the wavefront was then reconstructed using Zernike polynomial fitting. The results are summarized in Fig. 4c. The left panel shows the reconstructed Zernike coefficients for different aberration modes, where the measured values closely match the corresponding reference coefficients, confirming accurate phase retrieval. The right panel presents representative reconstructed wavefront profiles, comparing negative dielectric anisotropy LC droplets, positive dielectric anisotropy LC droplets, and simulation results.

Compared with positive dielectric anisotropy LC droplets, the negative dielectric anisotropy LC droplets exhibit improved reconstruction fidelity, with reduced deviations and more consistent recovery across multiple modes. Representative reconstructed wavefront profiles further confirm accurate recovery of both the spatial structure and magnitude of the imposed aberrations.

Together, the three panels in Fig. 4 demonstrate that intensity, wavelength, polarization, and phase information can be independently encoded and accurately retrieved using the same droplet platform under different operating conditions. Spectral information is encoded through wavelength-dependent intensity patterns, while polarization and phase information are retrieved through birefringence-based modulation and microlens-based wavefront sensing, respectively. This unified approach enables direct access to spectral, polarization, and phase information within a single device, without requiring separate optical components or complex system integration. These results demonstrate that printed negative dielectric anisotropy nematic LC droplets provide a compact and versatile platform for adaptive full-vectorial optical field sensing.

## Conclusion and discussion

In this work, we propose and demonstrate that inkjet-printed negative dielectric anisotropy nematic LC droplets provide an excellent candidate for an adaptive soft-matter photonic platform that integrates complex beam generation and full-vectorial optical field sensing within a single printed architecture. Through direct observations of the droplet optical response, we first reveal the multimodal optical functionality arising from the interplay between droplet geometry and electrically tunable birefringence. Building upon these properties, we demonstrate adaptive complex beam manipulation with two key capabilities: encompassing complex beam generation through voltage-tunable skyrmionic optical fields and adaptive full-vectorial field sensing, where the same droplet array retrieves spectral, polarization, and phase information of incident light.

Compared with recently reported positive dielectric anisotropy nematic LC droplet demonstrations, negative dielectric anisotropy nematic LC droplets provide several important advantages. While positive dielectric anisotropy LC droplets have demonstrated polarization and phase sensing capabilities, their optical response is largely fixed once the droplet geometry is defined, limiting their ability to encode additional physical dimensions of the optical field without the use of alternative alignment layer/electrode configurations. In particular, although multi-wavelength operation can be supported in positive dielectric anisotropy LC droplets, we did not observe comparable wavelength-dependent optical signatures in their POM responses under passive states or in-plane electric fields driving conditions. This may be primarily due to the insufficient birefringent retardance achievable in such systems, which limits the formation of strong wavelength-dependent interference patterns. As a result, the transmitted intensity distributions remain weakly dependent on wavelength and cannot provide a robust intrinsic spectral signature for reliable wavelength discrimination. In contrast, the negative dielectric anisotropy LC droplets presented here exhibit a pronounced wavelength-dependent birefringent response, enabling direct spectral encoding and, to our knowledge, providing the first realization of a droplet-based spectral sensing prototype.

In addition to enabling spectral sensing, the negative dielectric anisotropy nematic LC droplets also improve the performance of polarization and phase measurements. For polarization sensing, the electric field-driven birefringence enhances modulation depth while maintaining robustness against variations in droplet geometry, leading to improved reconstruction accuracy compared with positive dielectric anisotropy nematic LC systems. For wavefront sensing, the improved reconstruction fidelity originates from the flatter droplet geometry and smaller overall structure of the negative dielectric anisotropy nematic LC platform, which produce cleaner focal spots and reduced centroiding errors. These geometric advantages also suppress the influence of both dynamic phase variation and geometric phase artefacts, resulting in more accurate and stable phase retrieval.

Several aspects of the present platform merit further investigation. In the skyrmionic field generation experiments, the apparent non-zero skyrmion number arises from the regularization of the reconstructed Stokes field rather than an unambiguous boundary-enforced topological change. Because the droplet centre and perimeter both exhibit vanishing retardance, the global mapping degree is naturally expected to remain close to a trivial boundary condition. Engineering boundary conditions, for example through laser-written birefringent structures[55], could help lift this limitation and enable droplets to support boundary-stabilized optical skyrmions. Beyond these aspects related to field generation, several sensing-related aspects of the present platform also merit further investigation. For wavefront sensing, although phase-related artefacts are significantly reduced, they are not fully eliminated. A potential hardware-level solution is to introduce orthogonal detection arms using a polarization beam splitter, allowing geometric phase components with opposite signs to be separated and cancelled. For spectral sensing, the current experiments are performed over the 460–619 nm range, primarily because the white LED source is unstable at other wavelengths, which makes reliable acquisition difficult. In principle, however, the droplet spectral response can be learned by a compact machine-learning model and extended to broader spectral ranges given sufficient training data. For polarization sensing, spectral and polarization measurements are currently demonstrated separately; however, with sufficiently rich training datasets spanning multiple wavelengths and polarization states, it should be possible to construct a unified inference model that simultaneously retrieves multiple parameters from a single droplet image, enabling multidimensional optical field characterization in a single measurement.

Overall, these results establish inkjet-printed negative dielectric anisotropy LC droplets as a versatile adaptive photonic platform capable of complex-beam generation and multidimensional optical field sensing within a compact and scalable architecture.

**Author Contributions**

J.G. and X.Q. contributed equally to this work.
C.H., J.G., and X.Q. conceived the experimental concept.
J.G. and X.Q. designed and performed the experiments, carried out data acquisition, and wrote the manuscript.
R.Z. assisted with data analysis and numerical simulations.
L.D. assisted with the construction of the spectral experimental setup.
Y.L. contributed to the development of machine-learning algorithms.
Z.Z. contributed to data processing and visualization tools for polarization analysis.
M.L., Q.H., J.M., and W.K. assisted with material fabrication and characterization.
Y.M. contributed to the development of simulation methods and machine-learning algorithms.
S.J.E. contributed to theoretical modelling and provided valuable discussions.
A.A.C.-P. provided resources and support for droplet fabrication.
C.H. and S.M.M supervised the project and reviewed all aspects of the work.
All authors discussed the results and contributed to the final manuscript.

**Acknowledgments**

AACP acknowledges funding from the NSF/CBET-EPSRC (Grant Nos. EP/W016036/1 and EP/S029966/1)

and the John Fell Fund via a Pump-Priming grant (0005176). C.H. acknowledges support from St John's College, the University of Oxford, and The Royal Society (URF/R1/241734). WK, SJE and SMM acknowledge the EPSRC-UK for a research grant (EP/W022567/1).

**Competing interests**

The authors declare no competing interests.

**Additional Information**

Correspondence and request for materials should be addressed to C.H. and S.M.M.

**Data Availability Statement**

The data that support the findings of this study are available in the Supporting Information of this article. Additional data are available from the corresponding author upon reasonable request.